\documentclass[aps,prc,twocolumn,superscriptaddress,preprintnumbers,nofootinbib,showkeys]{revtex4-2}

\usepackage[utf8]{inputenc}
\usepackage[english]{babel}
\usepackage{amssymb,amsthm,amsmath,amstext,amsbsy,amsopn}
\usepackage{bbm}
\usepackage{nicefrac}
\usepackage{slashed}
\usepackage{graphicx}
\usepackage{hyperref}
\usepackage{leftidx}
\usepackage{environ}
\usepackage{mathtools}
\usepackage{xspace}
\usepackage{array}
\usepackage{xcolor}
\usepackage{pgfplotstable}
\usepackage{isotope}

\newcommand{\bra}[1]{\langle #1|}

\newcommand{\ket}[1]{|#1\rangle}

\newcommand*\rvec[1]%
{\ensuremath{\overset{\smash{\raisebox{-1.5pt}{\tiny$\rightarrow$}}}{#1}}}
\newcommand*\lvec[1]%
{\ensuremath{\overset{\smash{\raisebox{-1.5pt}{\tiny$\leftarrow$}}}{#1}}}

\NewEnviron{subalign}[1][]{%
\begin{subequations}\begin{align}
  \BODY
\end{align}\label{#1}\end{subequations}
}

\NewEnviron{spliteq}{%
\begin{equation}\begin{split}
  \BODY
\end{split}\end{equation}
}

\newcolumntype{K}[1]{>{\centering\arraybackslash}p{#1}}

\def\lnl{$NN$+$3N\text{(lnl)}$}
\def\sat{NNLO$_{\text{sat}}$}
\def\dgo{$\Delta$NNLO$_{\text{GO}}$}

\newcommand{\ac}[1]{a^{\dagger}_{#1}}

\begin{document}

\title{On the calculation and use of effective single-particle energies.\\
The example of the neutron $1d_{3/2}$-$1d_{5/2}$ splitting along $\text{N}=20$ isotones}

\author{V.~Som\`a}
\affiliation{IRFU, CEA, Universit\'e Paris-Saclay, 91191 Gif-sur-Yvette, France}

\author{T.~Duguet}
\affiliation{IRFU, CEA, Universit\'e Paris-Saclay, 91191 Gif-sur-Yvette, France}
\affiliation{KU Leuven, Instituut voor Kern- en Stralingsfysica, 3001 Leuven, Belgium}

\date{\today}

\begin{abstract}
The rich phenomenology of quantum many-body systems such as atomic nuclei is complex to interpret. 
Often, the behaviour  (e.g. evolution with the number of constituents) of measurable/observable quantities such as binding or excitation energies can be best understood on the basis of a simplified {\it picture} involving auxiliary quantities that are not observable, i.e. whose values vary with parameters that are {\it internal} to the theoretical construction (contrarily to measurable/observable quantities). 
While being useful, the simplified interpretation is thus theoretical-scheme-dependent. 
This applies, in particular, to the so-called single-nucleon shell structure based on auxiliary {\it effective single-particle energies} (ESPEs). 
In this context, the present work aims at (i) recalling the way to compute ESPEs out of solutions of many-body Schr\"odinger's equation, (ii) illustrating the use of ESPEs within the frame of state-of-the-art ab initio calculations to interpret the outcome of a recent nuclear experiment and (iii) demonstrating the impact of several alterations to the computation of ESPEs. 
While the chosen alterations constitute approximations within the ab initio scheme, they are built-in when employing other theoretical constructs at play in nuclear physics. 
The present considerations are thus meant to empirically illustrate variations that can be expected between ESPEs computed within different (equally valid) theoretical schemes.  
\end{abstract}

\maketitle

\section{Introduction}
\label{sec_intro}

Within the so-called ab initio theoretical scheme~\cite{Ekstrom23}, atomic nuclei are depicted as a collection of A interacting point-like protons and neutrons. 
Other, complementary, theoretical schemes employed to address the rich and diverse nuclear phenomenology are based on different frameworks and assumptions~\cite{Bender03, Caurier05}.
One paramount example is the valence-space shell model where only a subset of the A nucleons is active within a reduced space defined through a partitioning of the A-body Hilbert space and interact via adequate effective forces\footnote{Additional theoretical schemes exist and are based on yet different, e.g. collective, degrees of freedom.}. 
Each theoretical scheme comes with an (often implicit) associated scale, and amounts to solving of an adequate dynamical equation, out of which A-body observables are obtained.
For example, the ab initio scheme requires to solve an A-body Schr\"odinger equation on the full A-body Hilbert space yielding A-body eigen-energies and eigen-functions from which other nuclear observables can be computed.

In any (useful) theoretical scheme, the handling of many-body objects is complicated from several points of view. For instance, A-body wave functions are both difficult to compute and hard to represent in an intuitive way. In order to best deal with this complexity and interpret the associated phenomenology, it is natural to resort to simpler quantities that, by distilling some essential information, aid in our understanding of correlated many-body systems. This step, however, comes with fundamental limitations. While A-body observables can be compared without ambiguity to their experimental counterparts, simpler (reduced) quantities typically depend on the employed theoretical scheme, i.e. contrarily to measurable quantities their value depend on parameters that are only {\it internal} to the theoretical construction\footnote{This dependence does {\it not} relate to approximations but clearly separate observable and non-observable quantities.}. Because the value taken by these auxiliary quantities  depends on the employed theoretical scheme, the interpretation of A-body observables (which hopefully match experimental data) thus delivered is not absolute, i.e. different but equally valid interpretations of the same observables can be generated.

A prime example relates to the interpretation of A-body observables based on a collection of independent one-body orbitals\footnote{While the empirical relevance of such a picture is not foreign to the potential success of a so-called mean-field or independent-particle {\it approximation}, the present discussion relates to the possibility to extract such a picture from the {\it fully correlated} solution of the problem. The impact of potential approximations come on top of such a consideration.}.
This picture, inspired from atomic structure, can be accessed unambiguously through a well-defined theoretical procedure, first proposed by Baranger~\cite{Baranger70}, delivering effective single-nucleon states and energies out of the {\it correlated} A-body solution\footnote{The procedure is unambiguous and well-defined because it does not relate to any approximation and only invokes (ideally exact) solutions of the many-body problem. Furthermore, the resulting quantities are independent of the auxiliary single-particle basis used to expand the dynamical equation at play. See Sec.~\ref{sec_theory} for details.}. The resulting effective single-particle energies (ESPEs) are often used by nuclear structure practitioners to analyse and {\it interpret} in simple terms the evolution of nuclear properties across the Segr\`e chart.

In Refs.~\cite{Duguet12,Duguet15b} the non-observable character of ESPEs was demonstrated formally and illustrated computationally via actual ab initio many-body calculations. Once the scale and scheme dependence of ESPEs, i.e. of the single-nucleon shell structure, has been recognised, it is well justified to resort to it to interpret the nuclear phenomenology {\it within a fixed theoretical scheme}. While the formal definition of ESPEs is universal, the very characteristics of a given theoretical construction may lead to specific restrictions of the general definition and induce a scheme dependence. 

The present paper is of pedagogical nature and aims at illustrating such a feature. Employing the ab initio scheme based on self-consistent Green's function techniques~\cite{Soma20b}, the computation of ESPEs is first illustrated. Next, two alterations are performed to qualitatively illustrate differences that can be expected with ESPEs computed within other (equally valid) theoretical schemes. 
This study is accompanied by a detailed assessment of theoretical uncertainties, ensuring that present findings are not significantly affected by the various approximations at play in the many-body calculation.
The analysis is carried out focusing on a particular physics case of recent interest, i.e. the evolution of the splitting between neutron $\ell=2$ ESPE spin-orbit partners in $\text{N}=20$ isotones~\cite{Jongile23}. 
In particular, an observed reduction of this spin-orbit splitting going from $^{40}$Ca to $^{36}$S has been interpreted across several theoretical schemes as a fingerprint of the action of the tensor component of the two-nucleon interaction in atomic nuclei~\cite{Jongile23}.

The manuscript is organised as follows. In Sec.~\ref{sec_theory} Baranger ESPEs are formally introduced and their scheme (and scale) dependence is discussed in some detail.
Sections~\ref{sec_setting} and \ref{sec_physics} briefly present the employed theoretical scheme and the physics case under study.
In Sec.~\ref{sec_strength} results for the strength distributions are displayed.
The resulting ESPE spin-orbit splittings are shown in Sec.~\ref{sec_deltaSO}.
Finally, the sensitivity of the results to two alterations of the general definition of ESPEs is explored in Sec.~\ref{sec_sensitivity}. Section~\ref{sec_conclusions} ends the manuscript with some concluding remarks.

\section{Formal considerations}
\label{sec_theory}

\subsection{Theoretical scheme}
\label{subsec_theoretical scheme}

The ab initio theoretical scheme describe the atomic nucleus as a collection of A structure-less nucleons (Z protons and N neutrons) interacting via k-nucleon interactions derived within the frame of chiral effective field theory ($\chi$EFT)~\cite{Hammer20b}. The dynamical equation to be solved to access properties of stationary states is the time-independent A-body Schr\"odinger equation
\begin{equation}
H | \Psi^{\text{A}}_{\kappa} \rangle =E^{\text{A}}_{\kappa} | \Psi^{\text{A}}_{\kappa}\rangle \, \, ,
\label{Abodyschroe} 
\end{equation}
which delivers A-body eigenenergies $E^{\text{A}}_{\kappa}$ and eigenstates $| \Psi^{\text{A}}_{\kappa}\rangle$ starting from the Hamiltonian $H$ encoding inter-nucleon interactions. This Hamiltonian reads in a generic basis of the one-body Hilbert space  ${\cal H}_1$  as 
\begin{eqnarray}
\label{hamiltonian}
H &=&  T + V^{\text{2N}} + V^{\text{3N}} + \ldots  \nonumber  \\
&=&\sum_{pq} t_{pq} a^{\dagger}_p a_q \nonumber \\
&&+ \left(\frac{1}{2!}\right)^2 \sum_{pqrs} \overline{v}^{\text{2N}}_{pqrs} a^{\dagger}_p a^{\dagger}_q a_s a_r \nonumber \\
&&+ \left(\frac{1}{3!}\right)^2 \sum_{pqrstu} \overline{v}^{\text{3N}}_{pqrstu} a^{\dagger}_p a^{\dagger}_q a^{\dagger}_r a_u  a_t a_s \nonumber \\
&&+ \ldots\, , 
\end{eqnarray}
where $\overline{v}^{\text{2N}}_{pqrs}$ and $\overline{v}^{\text{3N}}_{pqrstu}$ represent antisymmetrised matrix elements of two- (2N) and three-nucleon (3N) forces, while dots refer to higher-body forces that are omitted in practice.
Many-body observables other than total energies can be accessed from the knowledge of the A-body wave function, be it in its ground state $| \Psi^{\text{A}}_{0}\rangle$ or in an excited state $| \Psi^{\text{A}}_{\kappa \neq 0}\rangle$.

\subsection{Spectral quantities}
\label{subsec_spectral}

The endeavour to map a single-particle shell structure out of many-body eigenstates primarily relates to one-nucleon addition and removal processes. 
Experimentally, those processes are typically achieved by transfer reactions, i.e. nuclear collisions where the projectile either donates a nucleon to the A-body target (\emph{stripping} reactions) or acquires a nucleon from the A-body target (\emph{pickup} reactions).
Such collisions leave the A+1 (A-1) system in an eigenstate $\ket {\Psi^{\text{A+1}}_{\mu}}$ ($\ket {\Psi^{\text{A-1}}_{\nu}}$) whose eigenenergy $E^{\text{A+1}}_{\mu}$ ($E^{\text{A-1}}_{\nu}$) is measured by the experimental apparatus. 

From the latter quantities one can construct one-nucleon addition and removal energies 
\begin{equation}
E^{\pm}_\lambda \equiv \pm \big( E^{\text{A} \pm 1}_\lambda - E^{\text{A}}_0 \big) \, ,
\label{sep_en}
\end{equation}
i.e. observable quantities that can be cleanly compared between experiment and theory. 
However, they do not qualify as single-particle energies defining a so-called single-nucleon shell structure\footnote{The simplest way to appreciate such a key statement is by realising that one-nucleon addition and removal energies cannot be put in bijection with a basis of the one-body Hilbert space. Indeed, starting from a one-body Hilbert space of dimension $n_{\text{dim}}$ and building  A-body Hilbert spaces via repeated tensor products of it, the cardinal of one-nucleon addition (removal) energies to (from) an A-body system scales as $n^{\text{A}+1}_{\text{dim}}$ ($n^{\text{A}-1}_{\text{dim}}$) whereas the number of any meaningful set of single-particle energies is necessarily $n_{\text{dim}}$.}. A specific procedure is thus required to go from one-nucleon separation energies to ESPEs that combines the knowledge of the former with information about the process of adding (removing) a nucleon to (from) the ground-state of the A-body system in (from) a specific single-particle state. An additional subtlety is that the single-particle states in question are not known \emph{a priori} and must {\it emerge} from the procedure along with associated ESPEs.

Information on one-nucleon transfer can be typically encoded in the probability amplitudes to reach the eigenstate $\ket {\Psi^{\text{A+1}}_{\mu}}$ ($\ket {\Psi^{\text{A-1}}_{\nu}}$) by adding (removing) a nucleon in (from) a one-body basis state $| p \rangle \equiv \ac {p} | 0 \rangle$ to (from) the ground state $\ket{\Psi^{\text{A}}_{0}}$, introduced according to
\begin{eqnarray}
U^{p}_{\mu} &\equiv& \bra {\Psi^{\text{A}}_{0}} a_p \ket {\Psi^{\text{A+1}}_{\mu}} \,\, \,  \left(V^{p}_{\nu} \equiv \bra {\Psi^{\text{A}}_{0}} a^\dagger_p \ket {\Psi^{\text{A-1}}_{\nu}} \right) \label{eq:defu}
\end{eqnarray}
and collected in the vector\footnote{Bold symbols denote tensors in the one-body Hilbert space.} $\mathbf{U}_{\mu}$ ($\mathbf{V}_{\nu}$). 
From those, spectroscopic probability matrices for one-nucleon addition and removal processes $\mathbf{S}_{\mu}^{+}\equiv \mathbf{U}_{\mu} \mathbf{U}^{\dagger}_{\mu}$ and $\mathbf{S}_{\nu}^{-}\equiv (\mathbf{V}_{\nu}\mathbf{V}^{\dagger}_{\nu})^{\ast}$ are defined. 
Their elements read as\footnote{Spectroscopic factors are obtained by tracing spectroscopic probability matrices over the one-body Hilbert space ${\cal H}_1$
\begin{subequations}
\label{spectrofactor}
\begin{eqnarray}
SF_{\mu}^{+} &\equiv& \text{Tr}_{{\cal H}_{1}}\!\left[ \mathbf{S}_{\mu}^{+}\right] =  \sum_{p \in {\cal H}_{1}} \left|U^{p}_{\mu}\right|^2 \, \, , \nonumber \\
SF_{\nu}^{-} &\equiv& \text{Tr}_{{\cal H}_{1}}\!\left[\mathbf{S}_{\nu}^{-} \right] = \sum_ {p \in {\cal H}_{1}} \left|V^{p}_{\nu}\right|^2  \nonumber \,\, .
\end{eqnarray}
\end{subequations}
A spectroscopic factor delivers the total probability that an eigenstate $\ket {\Psi^{\text{A+1}}_{\mu}}$ ($\ket {\Psi^{\text{A-1}}_{\nu}}$) of the A+1 (A-1) system can be described as a nucleon added to (removed from) a single-particle state on top of the ground state of the A-nucleon system. While being scale and scheme dependent~\cite{jennings11a}, spectroscopic factors are independent of the one-body basis employed to expand spectroscopic matrices.}
\begin{subequations}
\label{spectroproba}
\begin{eqnarray}
S_{\mu}^{+pq} &\equiv&  \bra {\Psi^{\text{A}}_{0}} a_p \ket {\Psi^{\text{A+1}}_{\mu}} \bra {\Psi^{\text{A+1}}_{\mu}} a^\dagger_q \ket {\Psi^{\text{A}}_{0}}    \, \, \, , \label{spectroprobaplus} \\
S_{\nu}^{-pq} &\equiv& \bra {\Psi^{\text{A}}_{0}} a^\dagger_q \ket {\Psi^{\text{A-1}}_{\nu}} \bra {\Psi^{\text{A-1}}_{\nu}} a_p \ket {\Psi^{\text{A}}_{0}}    \, \, \, . \label{spectroprobamoins}
\end{eqnarray}
\end{subequations}

The spectroscopic information in Eqs.~\eqref{sep_en} and \eqref{spectroproba} is assembled into the {\it spectral function}, an energy-dependent matrix defined on ${\cal H}_1$, according to
\begin{eqnarray}
\mathbf{S}(\omega) &\equiv& \mathbf{S}^{+}(\omega) + \mathbf{S}^{-}(\omega) \nonumber \\
 &\equiv& \!\!\!\!\! \sum_{\mu \in {\cal H}_{A\!+\!1}} \!\!\! \mathbf{S}_{\mu}^{+} \,\, \delta(\omega -E_{\mu}^{+}) +  \!\!\!\!\!\sum_{\nu\in {\cal H}_{A\!-\!1}} \!\!\! \mathbf{S}_{\nu}^{-}  \,\, \delta(\omega -E_{\nu}^{-}) , \label{spectralfunction}
\end{eqnarray}
where the first (second) sum is restricted to eigenstates of $H$ in the Hilbert space ${\cal H}_{\text{A}\!+\!1}$ (${\cal H}_{\text{A}\!-\!1}$) associated with the A+1 (A-1) system.

\subsection{Baranger one-body Hamiltonian}
\label{subsec_baranger}

Next, it is useful to introduce the moments of the spectral function, computed according to
\begin{equation}
\mathbf{M}^{(n)} \equiv \int_{-\infty}^{+\infty} \omega^{n} \, \mathbf{S}(\omega) \, d\omega  \label{spec_funct_moments}
\end{equation}
and constituting energy-independent matrices on ${\cal H}_1$. 

The zeroth moment can be shown to be the identity matrix\footnote{This result reflects the anti-commutation properties of fermionic creation and annihilation operators. As such, it is scale and scheme independent as well as independent of the one-body basis used to expand spectroscopic matrices.}
\begin{equation}
\mathbf{M}^{(0)}  = \sum_{\mu\in {\cal H}_{A\!+\!1}} \mathbf{S}_{\mu}^{+} + \sum_{\nu\in {\cal H}_{A\!-\!1}} \mathbf{S}_{\nu}^{-} = \mathbf{1} \, . \label{normalizationspectro}
\end{equation}
This sum rule provides each diagonal matrix element of $\mathbf{S}(\omega)$ with the meaning of a probability distribution function in the statistical sense, i.e., the combined probability of adding a nucleon to or removing a nucleon from a specific single-particle basis state $| p \rangle$ integrates to 1 when summing over all final states of A$\pm$1 systems. 

The first moment of the spectral function defines the so-called one-body Baranger, or centroid, Hamiltonian
\begin{align}
\mathbf{M}^{(1)}  
&= 
\sum_{\mu\in {\cal H}_{A\!+\!1}} \mathbf{S}_{\mu}^{+} E_{\mu}^{+} + \sum_{\nu\in {\cal H}_{A\!-\!1}}  \mathbf{S}_{\nu}^{-} E_{\nu}^{-} \nonumber \\
&\equiv  \mathbf{h}^{\text{cent}}  \label{defsumrule} \, .
\end{align}

\subsection{Effective single-particle energies}
\label{subsec_ESPE}

Effective single-particle energies are nothing but the {\it eigenvalues} of  $\mathbf{h}^{\text{cent}} $~\cite{French66, Baranger70}, i.e. they are obtained by solving the one-body eigenvalue problem
\begin{eqnarray}
\mathbf{h}^{\text{cent}} \, | \psi^{\text{cent}}_b \rangle &=& e^{\text{cent}}_{b} \, | \psi^{\text{cent}}_b \rangle \, , \label{HFfield3}
\end{eqnarray}
and are thus independent of the one-body basis used to expand $\mathbf{h}^{\text{cent}}$ in matrix form. In fact, solving Eq.~\eqref{HFfield3} does not only provide ESPEs but also delivers Baranger one-body eigenstates the nucleon is effectively added to or removed from.  The associated basis of ${\cal H}_1$ is denoted as $\{c^\dagger_{b}\}$ and is also independent of the one-body basis initially used to expand spectroscopic probability matrices. 

Focusing as an example on an even-even nucleus, the $J^{\pi} = 0^+$ character of its ground state $\ket{\Psi^{\text{A}}_{0}}$ and the rotational invariance of the system make $\mathbf{h}^{\text{cent}}$ to be spherically symmetric. As a result, Baranger basis states carry spherical quantum numbers $b\equiv(n_b,\pi_b,j_b,m_b,\tau_b)$ denoting respectively the principal quantum number, the parity, the total angular momentum and its projection on the, e.g., z-axis as well as the isospin projection.

Employing the Baranger basis, an ESPE involves diagonal spectroscopic probabilities
\begin{eqnarray}
e^{\text{cent}}_{b} &\equiv&  \sum_{\mu \in {\cal H}_{A\!+\!1}} S_{\mu}^{+bb} E_{\mu}^{+} + \sum_{\nu\in {\cal H}_{A\!-\!1}}  S_{\nu}^{-bb} E_{\nu}^{-}  \, , \label{HFfield2}
\end{eqnarray}
and appears to be nothing but a centroid, i.e., the arithmetic average of one-nucleon separation energies weighted by the probability to reach the corresponding A+1 (A-1) eigenstates by adding (removing) a nucleon to (from) the single-particle state $| \psi^{\text{cent}}_b \rangle$. 

\subsection{Discussion}
\label{subsec_discussion}

\subsubsection{Scale and scheme dependence}
\label{subsec_remarks}

Equations~\eqref{defsumrule}-\eqref{HFfield2} provide a model-independent {\it definition} of ESPEs, in the sense that these equations are applicable to any theoretical scheme delivering eigenstates and eigenenergies of A and A$\pm$1 systems.  Nevertheless, contrarily to one-nucleon separation energies (Eq.~\eqref{sep_en}) or other measurable quantities, ESPEs are non-observable  due to the dependence of spectroscopic matrices on i) the employed theoretical scheme and ii) the resolution scale associated with the Hamiltonian.
\begin{enumerate}
\item[i)]
It can indeed be proven both formally and numerically that spectroscopic matrices depend on the resolution scale of the Hamiltonian~\cite{Duguet15b}, which in practice can be varied via unitary transformations\footnote{The independence of observable/measurable quantities under  unitary transformations is a testimony of the {\it surplus structure} entering the mathematical formulation of physical theories such as quantum mechanics or quantum field theories~\cite{redhead03a}. Conversely, this mathematical surplus precisely reflects onto {\it non}-observable quantities such as ESPEs that depend on the arbitrary value chosen to fix this surplus in any practical computation.}. By definition, ESPEs inherit such a dependence while true observables remain invariant.
\item[ii)]
The theoretical-scheme dependence can be exemplified by considering the shell-model scheme based on a valence space built out of a single major spherical harmonic oscillator shell. In such a model, with a single one-body basis state per angular momentum value in the active space, spectroscopic matrices are diagonal in the employed spherical harmonic oscillator basis {\it by construction}, and so is $\mathbf{h}^{\text{cent}}$. The Baranger basis is thus bound to be the harmonic oscillator basis and the diagonalisation of $\mathbf{h}^{\text{cent}}$ has no impact on the ESPE values. Contrarily, in the ab initio scheme all spherical harmonic oscillator basis states are active  such that  spectroscopic probability matrices are only {\it block} diagonal with respect to angular momentum. The diagonalisation of $\mathbf{h}^{\text{cent}}$ leads to a mixing of the principle quantum numbers, resulting into a Baranger basis that differs explicitly from the initial spherical harmonic oscillator basis used to expand the problem. Under the hypothesis that both theoretical schemes provide the same observable one-nucleon separation energies\footnote{Both reproducing experimental values within a desired accuracy.}, the different spectroscopic probability matrices and the non-trivial diagonalisation of $\mathbf{h}^{\text{cent}}$ at play in the ab initio scheme lead to different ESPEs.
\end{enumerate}
Eventually, the nucleonic shell structure is intrinsically non observable such that ESPEs computed from different theoretical scales and/or schemes should not be compared. Even though in practice this (usually implicit) dependence often leads to mildly different and thus qualitatively similar ESPEs, the opposite can also be true in certain cases~\cite{Duguet15b}. 

\subsubsection{Direct-reaction scheme}
\label{subsec_directreaction}

The above considerations also apply to ESPEs that are usually said to be ``extracted" from experimental data. While this ``extraction" is typically done on the basis of one-nucleon transfer reaction experiments, it involves a theoretical model to go from experimental many-body cross sections to spectroscopic probability matrices. To do so, the direct-reaction cross section is postulated to factorize into a single-particle cross section and a diagonal element of the spectral probability matrix\footnote{In this theoretical scheme, spectroscopic probability matrices are assumed {\it by construction} to be diagonal in the one-body basis used to compute the single-particle cross section. While this assumption resembles the one at play in shell-model structure calculation based on one major spherical harmonic oscillator shell (as discussed above), the two are independent from each other.}. Computing the former via a reaction model allows one to extract the latter from the experimental direct-reaction cross section. Combined with experimental one-nucleon separation energies, ESPEs are then computed via Eq.~\eqref{HFfield2}.

In spite of using experimental one-nucleon separation energies, the necessity to employ a theoretical model to calculate (diagonal) spectroscopic probability matrices that {\it cannot} be measured makes ESPEs as theoretical as those entirely computed from a structure model\footnote{Structure theoretical schemes all rely on experimental data at the time of their construction, e.g. ab initio theoretical schemes rely on nuclear Hamiltonians whose low-energy constants are fitted to a set of experimental data.}. Consequently, the ``extracted" ESPEs are as scale and scheme dependent as before. As a matter of fact, the present discussion makes clear that the terminology ``extracted" or ``reconstructed" from experiment is fundamentally improper. Indeed it implies that ESPEs with unique values leave their fingerprint in experimental data such that a meaningful objective is to reveal those values via some ``extraction" procedure. This is simply not the case. In the end, the above procedure is referred to below as the ``direct-reaction" approach to ESPEs to differentiate it from those obtained based on structure calculations (also of various types as discussed above).

The strategy to rely on experimental separation energies and many-body cross sections comes with a limited accessible energy range such that the sums entering Eq.~\eqref{defsumrule} are truncated accordingly. While one would typically wish to access all final states associated with non-zero one-neutron removal/additional cross sections, it is not strictly possible in practice. The corresponding accessible range can either be seen as an approximation to a hypothetical more complete direct-reaction scheme or as a part of its actual {\it definition}\footnote{Other schemes, e.g. the valence-space shell model or the ab initio scheme, also comes with a (often left implicit) limited range of accessible final states.}. Either way, the accessible range needs (i) to at least incorporate the final states carrying the largest cross-sections for the delivered picture to be empirically satisfying and (ii) to be similar in different nuclei for the evolution of that picture to be meaningful.

\subsubsection{Consistency}
\label{subsec_consistency}

The above considerations come before more pedestrian ones associated with phenomenological adjustments and inconsistencies entering the reaction model, e.g. the different scales/schemes used to compute the target density and the nucleon-target optical potential etc. While the theoretical scale and scheme dependence of ESPEs is inevitable, and thus not a problem per se once it is properly appreciated, inconsistencies are problematic because they forbid to properly control such a scale and scheme dependence. In this respect, the development of consistent approaches to nuclear structure and reactions is crucial~\cite{Hebborn23}.  

\section{Ab initio structure calculations}
\label{sec_results}

\subsection{Computational setting}
\label{sec_setting}

The present study is based on ab initio self-consistent Gorkov-Green's function (GGF) calculations~\cite{Soma11,Soma14a}.
The GGF approach is a full-space correlation-expansion method suited to the description of semi-magic mid-mass nuclei (see Ref.~\cite{Soma20b} for some recent applications).
This technique focuses on the computation of the one-nucleon propagator within the correlated A-body system. 
In its Lehmann representation, the propagator provides direct access to spectroscopic probability matrices (Eq.~\eqref{spectroproba}) and associated one-nucleon separation energies (Eq.~\eqref{sep_en}).
As a result, ESPEs can be easily computed by diagonalising the first moment of the spectral function, Eq.~\eqref{defsumrule}. 
Alternatively, one can evaluate the so-called static one-nucleon self-energy using correlated ground-state density matrices, which provides an equivalent expression for the centroid Hamiltonian defined in Eq.~\eqref{HFfield3}~\cite{Duguet15b}.

Gorkov equations are solved at second order in the Algebraic Diagrammatic Construction [ADC(2)]~\cite{Soma11}, which supplement the first-order Hartree-Fock-Bogolyubov (HFB) mean-field approximation with dominant many-body correlations.
The ADC(2) approximation provides most of the correlation energy beyond HFB and generates the fragmentation of the single-particle strength.
The convergence of theoretical quantities of interest with respect to the many-body truncation is assessed at times by using results at third-order, i.e. ADC(3), from Refs.~\cite{Duguet17b, Soma20a}. 

Different sets of two- (2N) plus three-nucleon (3N) interactions based on $\chi$EFT are employed in this work.
Most of the calculations are performed with the \sat{} Hamiltonian~\cite{Ekstrom15}, which is characterised by a simultaneous fit of all (2N+3N) coupling constants on binding energies and charge radii of selected carbon and oxygen isotopes, in addition to two-, three- and four-nucleon observables.
In order to explore the sensitivity of the results to the employed interaction, other Hamiltonians have also been used.
The \lnl{} Hamiltonian~\cite{Soma20a} features low-energy constants that are fitted on few-body observables only.
In addition, it is evolved to a low-resolution scale $\lambda = 2~\text{fm}^{-1}$ via a similarity renormalisation group (SRG) unitary transformation.
The \dgo{} Hamiltonian~\cite{Jiang20} explicitly includes $\Delta$-isobar degrees of freedom and is fitted on few-body observables as well as nuclear matter.
Two versions with different momentum cutoffs have been tested, namely $\Lambda = 394$ and 450 MeV.
The corresponding interactions are denoted with \dgo(394) and \dgo(450) respectively.

Gorkov equations are expanded using a one-body spherical harmonic oscillator basis whose dimension $n_{\text{dim}}$ is controlled by the parameter $e_\text{max} \equiv \text{max} (2n+\ell)$, where $n$ and $\ell$ represent the principal quantum number and orbital angular momentum of the basis eigenfunction, respectively.
Unless stated otherwise, $e_\text{max}=13$ is set in the following.
Three-body matrix elements are additionally truncated to $e_\text{3max} =16 < 3 \, e_\text{max}$ due to memory constraints.
These values of basis parameters are typically sufficient to yield converged calculations of nuclei with mass $A \lesssim 60$.

Finally, three-body forces are included by averaging 3N operators over one nucleon, thus obtaining an effective 2N interaction.
This average is performed at each iteration step by convoluting with the fully correlated one-body ground-state density matrix, which goes beyond the standard normal-ordered two-body approximation~\cite{Carbone13}.

\subsection{Physics case}
\label{sec_physics}

Present calculations were motivated by a recent experimental campaign carried out at iThemba LABS~\cite{Jongile23}. 
In this experiment, the $^{36}\text{S}(p, d)^{35}\text{S}$ transfer reaction was studied at proton energy $E_p=66~\text{MeV}$, yielding information on 98 final states in $^{35}\text{S}$ up to an excitation energy of 16 MeV.
Measurements at several scattering angles allowed the extraction of the angular distribution of the cross section for many of these states, from which the transferred angular momentum could be determined. As a result, refined spectroscopic data were collected in $^{35}\text{S}$. Combining such information with that from the $^{36}\text{S}(d, p)^{37}\text{S}$ reaction~\cite{Cameron12}, neutron $1d_{5/2}$ and $1d_{3/2}$ ESPEs can be calculated in $^{36}\text{S}$ using the direct-reaction scheme discussed above based on the reaction model plus experimental one-nucleon separation energies~\cite{Jongile23}.

The interest in the $^{36}\text{S}$ nucleus lies in the fact that the splitting between neutron $1d_{5/2}$ and $1d_{3/2}$ ESPE spin-orbit partners displays an anomalously low value, in contrast with what is expected from a well-understood systematic trend~\cite{Mairle93}.
Performing an analogous\footnote{Using the same direct-reaction framework, which minimises the bias introduced by eventual inconsistencies entering the analysis of experimental data.} analysis on existing experimental data~\cite{Matoba93, Nesaraja16}, the corresponding splitting can be consistently calculated in  $^{40}\text{Ca}$ as well. Doing so, one finds that the energy splitting based on the direct-reaction approach \emph{increases} from $^{36}\text{S}$ to $^{40}\text{Ca}$, in clear contrast with the global trend that instead predicts a \emph{reduction}~\cite{Jongile23}. While the latter is interpreted based on the spin-orbit component of the nuclear interaction~\cite{Mairle93}, the analysis performed in Ref.~\cite{Jongile23} based on several theoretical structure models, including the presently used ab initio scheme, interpreted the anomaly in connection with the tensor force at play in atomic nuclei. To corroborate this interpretation, it will be interesting in the future to extend the direct-reaction analysis to the $1d_{3/2}-1d_{5/2}$ ESPE spin-orbit splitting to other $\text{N}=20$ isotones, in particular $^{38}\text{Ar}$ sitting between sulfur and calcium, and $^{34}\text{Si}$ with two protons less than $^{36}\text{S}$.

\subsection{Spectral function}
\label{sec_strength}

Focusing on the physics case described above, ab initio GGF-ADC(2) calculations of the even-even $\text{N}=20$ isotones $^{34}\text{Si}$, $^{36}\text{S}$, $^{38}\text{Ar}$ and $^{40}\text{Ca}$ are performed employing the \sat{} Hamiltonian. Following the procedure described in Sec.~\ref{sec_theory}, the $1d_{5/2}$ ($1d_{3/2}$) Baranger's ESPE is obtained by diagonalising the centroid one-body Hamiltonian (Eq.~\eqref{defsumrule}). 
Given that one-neutron addition and removal processes take place on a $J^\pi=0^+$ A-body ground-state, the matrix elements of $\mathbf{h}^{\text{cent}}$ in the $d_{5/2}$ ($d_{3/2}$) sub-block only involve $J^{\pi}=5/2^+$ ($J^{\pi}=3/2^+$) final states in the $\text{A}\pm1$ systems. 

\begin{figure}
\centering
\includegraphics[width=1.0\columnwidth]{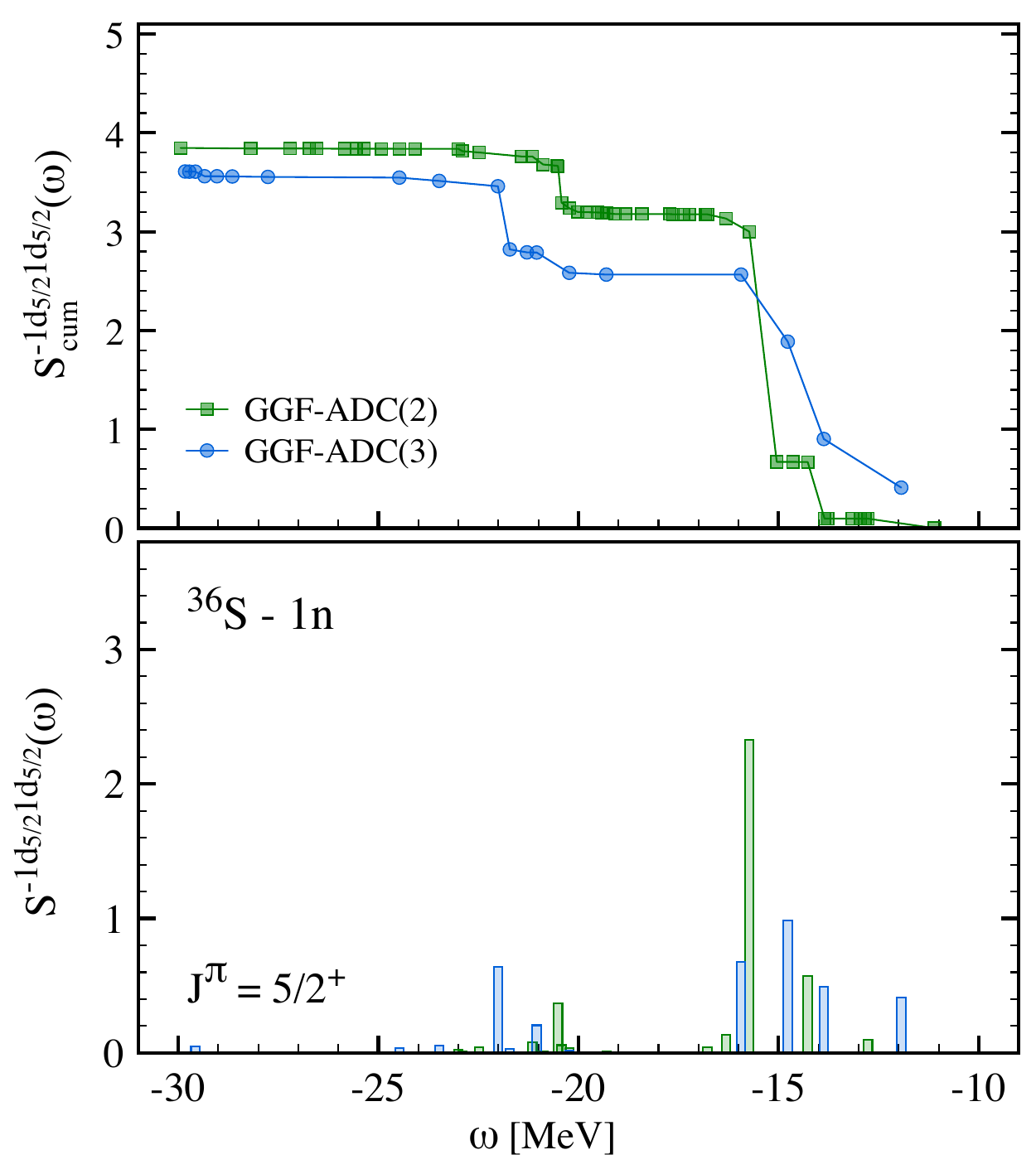}
\caption{(Bottom) Diagonal $1d_{5/2}$ part of the spectral function for one-neutron removal from the $^{36}$S ground state to $J^\pi = 5/2^+$ final states in $^{35}$S.
(Top) Corresponding cumulative strength.
Results from GGF-ADC(2) and GGF-ADC(3) calculations (the latter taken from Ref.~\cite{Duguet17b}) are shown. According to Eq.~\eqref{sep_en}, the x-axis scans (negative) one-neutron separation energies to $J^\pi = 5/2^+$ final states in $^{35}$S relative to the $J^\pi = 0^+$ $^{36}$S ground state. The computed separation energy to the $^{35}$S $J^\pi = 3/2^+$ ground-state is $E_0^- = -10.32$\, MeV ($-10.11$\, MeV)  in GGF-ADC(2) (GGF-ADC(3)).
}
\label{fig_strengthS}
\end{figure}

\begin{figure}
\centering
\includegraphics[width=1.0\columnwidth]{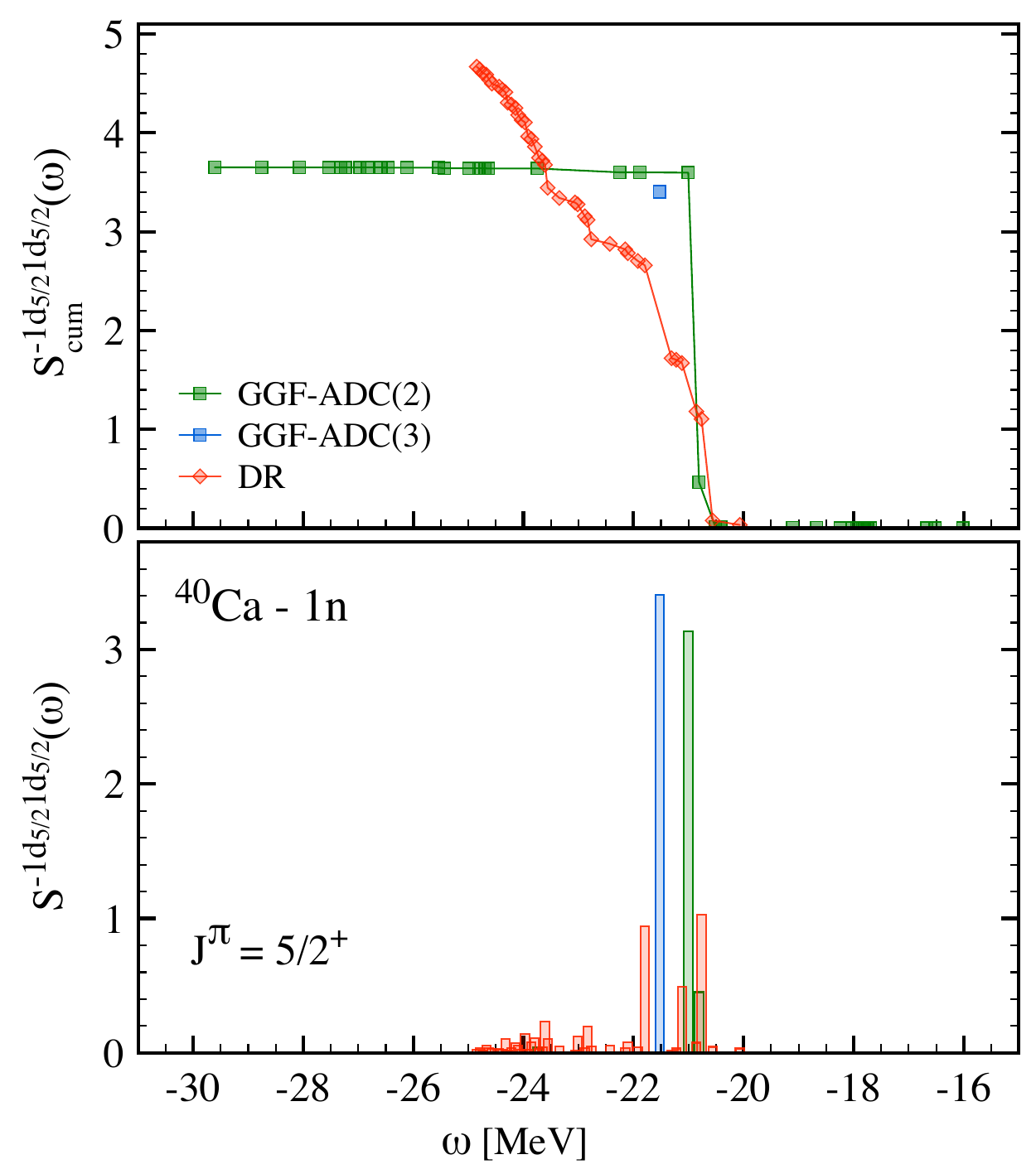}
\caption{(Bottom) Same as Fig.~\ref{fig_strengthS} for one-neutron removal from $^{40}$Ca.
Results from GGF-ADC(2) and GGF-ADC(3) calculations (the latter taken from Ref.~\cite{Soma20a}) are compared to the strength obtained via the direct-reaction approach~\cite{Matoba93}. The computed separation energy to the $^{39}$Ca $J^\pi = 3/2^+$ ground-state is $E_0^- = -13.84$\, MeV ($-14.03$\, MeV) in GGF-ADC(2) (GGF-ADC(3)).
}
\label{fig_strengthCa}
\end{figure}

Before coming to ESPEs, the bottom panel of Fig.~\ref{fig_strengthS} (\ref{fig_strengthCa}) displays the diagonal matrix element $S^{-  1d_{5/2}1d_{5/2}}(\omega)$ in the spherical harmonic oscillator basis of the removal part of the  GGF-ADC(2) spectral function in $^{36}$S ($^{40}$Ca). According to Eq.~\eqref{sep_en}, peaks (may) appear at (negative) one-neutron separation energies $E_{\mu}^-$ associated with $J^\pi = 5/2^+$ final states in $^{35}$S ($^{39}$Ca) calculated relative to the $J^\pi = 0^+$ $^{36}$S ($^{40}$Ca) ground state. More refined GGF-ADC(3) results taken from Refs.~\cite{Duguet17b} (\cite{Soma20a}) are also displayed for comparison.

In $^{36}\text{S}$, the diagonal $1d_{5/2}$ matrix element is well fragmented with peaks visible (in linear scale) down to about $-25$\,MeV (i.e. up to about $15$\,MeV excitation energy in $^{35}$S), the degree of fragmentation increasing going from GGF-ADC(2) to GGF-ADC(3), i.e. when including more correlations in the many-body treatment. The main peaks are found between $-12$ and $-16$ MeV. The situation is qualitatively different in $^{40}\text{Ca}$ where most of the strength is concentrated in one peak at about $-21$\,MeV (i.e. about $7$\,MeV excitation energy). Going to GFF-ADC(3) does not change the picture even though the main peak is shifted down by $0.5$\,MeV, which is partly due to the enlargement of the separation energy to the $3/2^+$ ground state.
The bottom panel of Fig.~\ref{fig_strengthCa} also displays the strength distribution obtained via the direct-reaction approach\footnote{By construction of the direct-reaction approach, for each $5/2^+$ final state, this is the only matrix element entering the $d_{5/2}$ sub-block of $\mathbf{h}^{\text{cent}}$, which is thus diagonal to begin with.}. The peaks position denotes in this case experimental one-neutron removal energies that can be compared to the theoretical ones visible in the GGF strength whenever the height of those peaks are non-zero in both schemes. Contrarily, the height of the peaks, i.e. the size of the matrix elements, is not observable and must thus be compared with care. Overall the removal strength computed within the direct-reaction approach is well fragmented in $^{40}\text{Ca}$ over the interval $\omega \in [-25,-20] $~MeV (i.e. between about 5 and 10\,MeV excitation energy in $^{39}\text{Ca}$), in contrast with the results from ab initio GGF structure calculations.

The qualitatively different behaviour in the two nuclei can be best appreciated in the top panels of Figs.~\ref{fig_strengthS} and \ref{fig_strengthCa} displaying the diagonal part of the cumulative one-neutron removal strength function defined as
\begin{eqnarray}
\mathbf{S}^{-}_\text{cum}(\omega) &\equiv& \int_{\omega}^{0} d\omega' \, \mathbf{S}^{-}(\omega') \, .
\end{eqnarray}
For a given spin-parity $j^\pi$, the asymptotic value $S^{-n\ell_jn\ell_j}_\text{cum}(-\infty)$ (reached at the extreme left of the figure) delivers the (non-observable) occupancy $n_{n\ell_j}$ of the corresponding harmonic oscillator basis state. This occupancy reaches the maximum (minimum) $2j+1$ (0) value for a fully occupied (empty) orbital in the independent-particle approximation, i.e. the departure from this value is a theoretical measure of many-body correlations\footnote{Strictly speaking, such an analysis must be conducted in the natural basis, i.e. using the eigenvalues of the one-body density matrix that are independent of the basis used to represent the one-body density operator.  While the present analysis based on single-particle occupations of harmonic oscillator basis states is useful to indicate the qualitative differences between nuclei and many-body truncations, it must only be seen as indicative.} in the ground state $| \Psi^{\text{A}}_{0} \rangle$ the neutron is removed from.


In the present case, the GGF-ADC(2) asymptotic value is similar for the two nuclei, i.e $n_{1d_{5/2}}\approx 3.8$ in $^{36}\text{S}$ and $n_{1d_{5/2}}\approx 3.6$ in $^{40}\text{Ca}$, to be compared to the maximum $2j+1=6$ value. This shows that many-body correlations deplete the orbit identically. However, the trajectory to reach this asymptotic value is very different in both nuclei, revealing the strength is fragmented over many more final states in $^{35}\text{S}$ than in $^{39}\text{Ca}$. Going from GGF-ADC(2) to GGF-ADC(3), additional many-body correlations further decreases $n_{1d_{5/2}}$ slightly by about 0.2 in both nuclei (i.e. rejecting the corresponding strength into the additional channel). However, whereas these additional correlations further fragment the strength in $^{36}\text{S}$, it does not do so in $^{40}\text{Ca}$ where it remains located in one peak.

In $^{36}\text{S}$, the fragmentation visible in the cumulative GGF-ADC(2,3) and direct-reaction removal spectral function are qualitatively similar across the whole energy range~\cite{Jongile23}. 
In $^{40}\text{Ca}$, conversely, the cumulative strengths are significantly different\footnote{By comparison, the diagonal $1d_{3/2}$ matrix element of the one-neutron removal strength to $J^\pi = 3/2^+$ final states is barely  fragmented in $^{36}\text{S}$ and $^{40}\text{Ca}$ for both the direct-reaction and the ab initio GGF ADC(2,3) approaches~\cite{Jongile23,Matoba93}. 
Most of the strength is concentrated  in the $3/2^+$ ground-state peaks. Contrarily, the $d_{5/2}$ strength function relates to excited $5/2^+$ states that typically leads to more fragmentation.}. 
On the one-hand, the GGF strength arguably suffers in $^{40}\text{Ca}$ from the incomplete character of the ADC(2,3) many-body approximations. 
This is aggravated by the use of the \sat{} interaction that tends to overestimate magic gaps and produce spectra that are too spread out, thus suppressing the onset of correlations beyond the mean field. 
This issue was shown to be particularly significant in doubly closed-shell nuclei~\cite{Soma20a}. On the other hand, the cumulative $1d_{5/2}$ strength from the direction-reaction is larger than in GGF-ADC(2) and does not seem to have reached a plateau within the experimentally accessed energy range.

\subsection{Spin-orbit splitting evolution}
\label{sec_deltaSO}

\begin{figure}
\centering
\includegraphics[width=1.0\columnwidth]{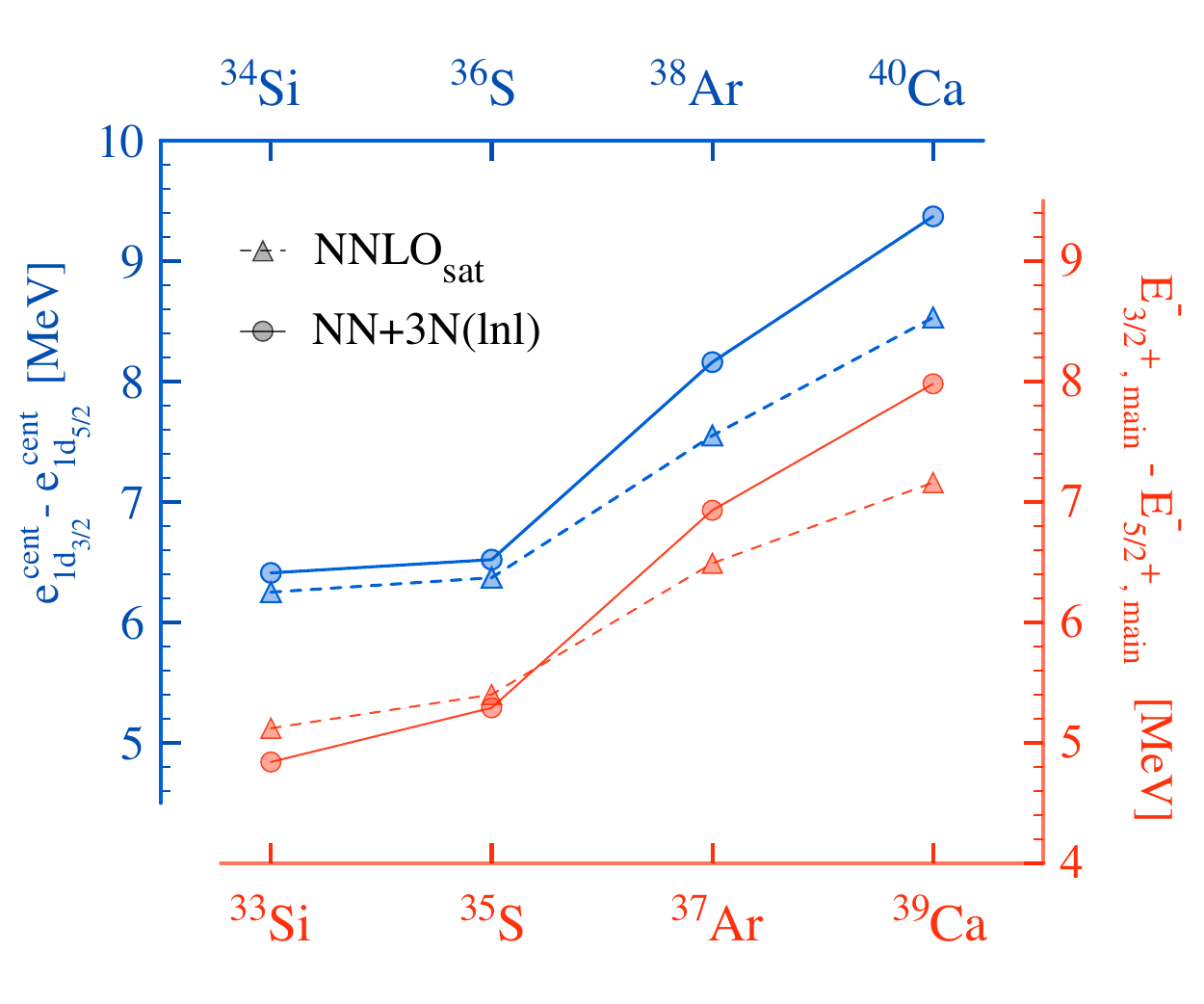}
\caption{Energy difference between the main $3/2^+$ and $5/2^+$ peaks (in red) compared to the difference of the resulting ESPEs (i.e., the spin-orbit splitting $\Delta^\text{SO}_{1d}$, in blue).
Calculations from GGF-ADC(2) are shown for the \sat{} and the \lnl{} Hamiltonians.
}
\label{fig_DeltaSO}
\end{figure}

Based on the spectral functions discussed above, Baranger ESPEs are computed. One-body spin-orbit splittings derive from the differences of two ESPEs associated with orbits characterised by the same principal quantum number $n$ and orbital angular momentum $\ell$ but the two possible total angular momenta $j_< = \ell -s$ and $j_> = \ell +s$, i.e.
\begin{equation}
\Delta^\text{SO}_{n\ell} \equiv e^{\text{cent}}_{n\ell_{j_<}} - e^{\text{cent}}_{n\ell_{j_>}}  \, ,
\label{eq_SO}
\end{equation}
with $s=1/2$ denoting the intrinsic spin. The splitting of present interest, i.e. the neutron $\Delta^\text{SO}_{1d} \equiv e^{\text{cent}}_{1d_{3/2}} - e^{\text{cent}}_{1d_{5/2}}$, is shown in Fig.~\ref{fig_DeltaSO} as a function of proton number.
Removing protons, one observes a steady decrease from $^{40}\text{Ca}$ down to $^{36}\text{S}$ followed by a stabilisation in $^{34}\text{Si}$.
Together with those produced with \sat, results obtained with the \lnl{} nuclear Hamiltonian are displayed. 
The two trends are similar, with the latter interaction producing a slightly larger splitting at $\text{Z}=20$. 
This difference presumably relates to the even worse overestimation of the $\text{N}=20$ magic gap by the \lnl{} Hamiltonian.

Figure~\ref{fig_DeltaSO} also displays the {\it many-body} spin-orbit splitting computed from the energies of the dominant\footnote{Carrying the largest spectroscopic factor.} $5/2^+$ and $3/2^+$ peaks in the one-neutron removal spectral functions. 
This many-body splitting is strictly observable and is intrinsically different from the one-body splitting $\Delta^\text{SO}_{1d}$ connecting {\it centroids} of the complete (diagonalised) spectral functions. 
As expected though, the ESPE spin-orbit splitting of hole-like states keeps a close memory of the many-body counterpart in the removal channel for both interactions, i.e. except for a shift up by 1-1.5 MeV, the trend is similar across the four isotones. The shift up is mainly due to the $5/2^+$ strength being more fragmented than the $3/2^+$ one.

\begin{figure*}
\centering
\includegraphics[width=0.99\columnwidth]{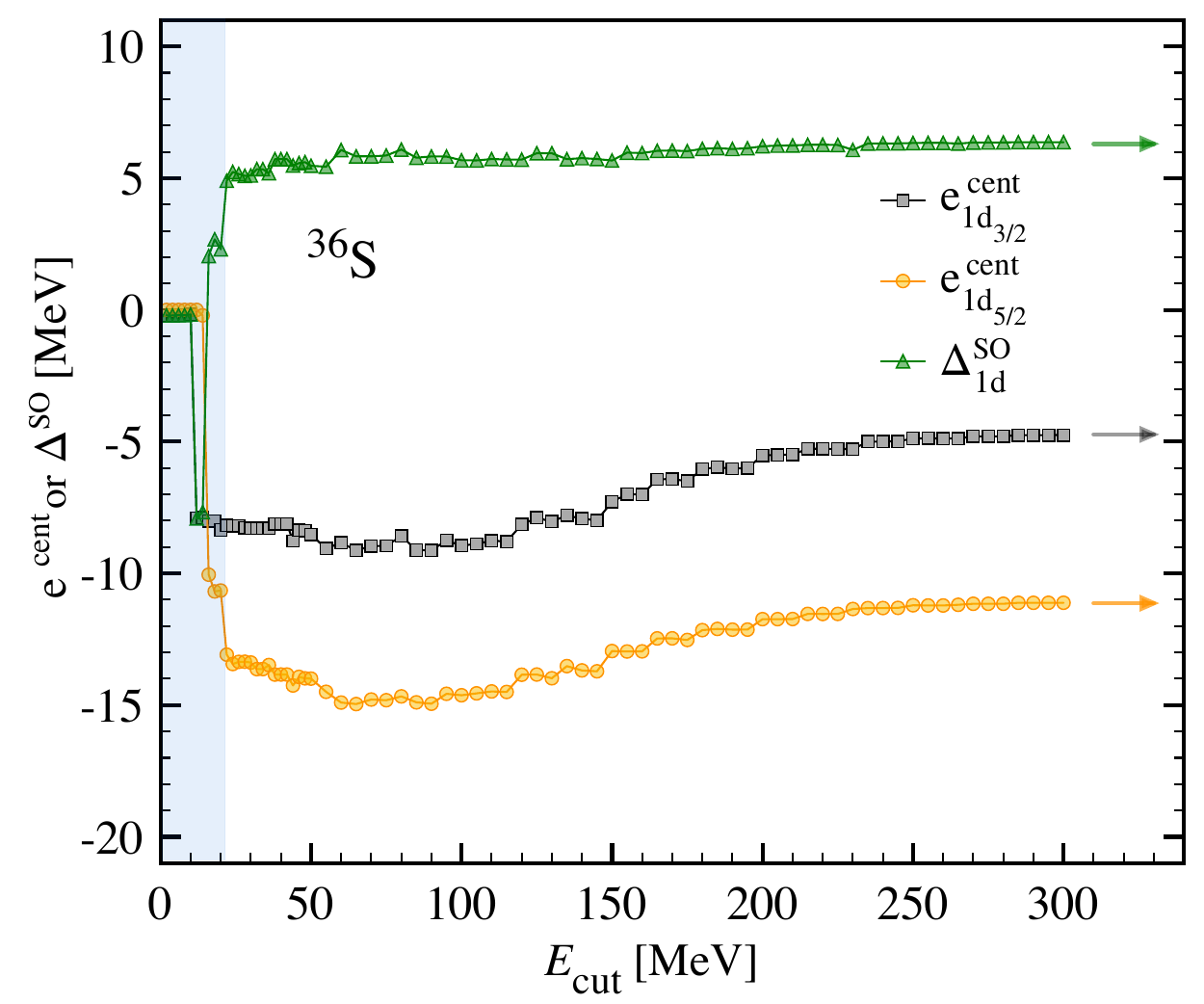}
\hspace{.3cm}
\includegraphics[width=0.99\columnwidth]{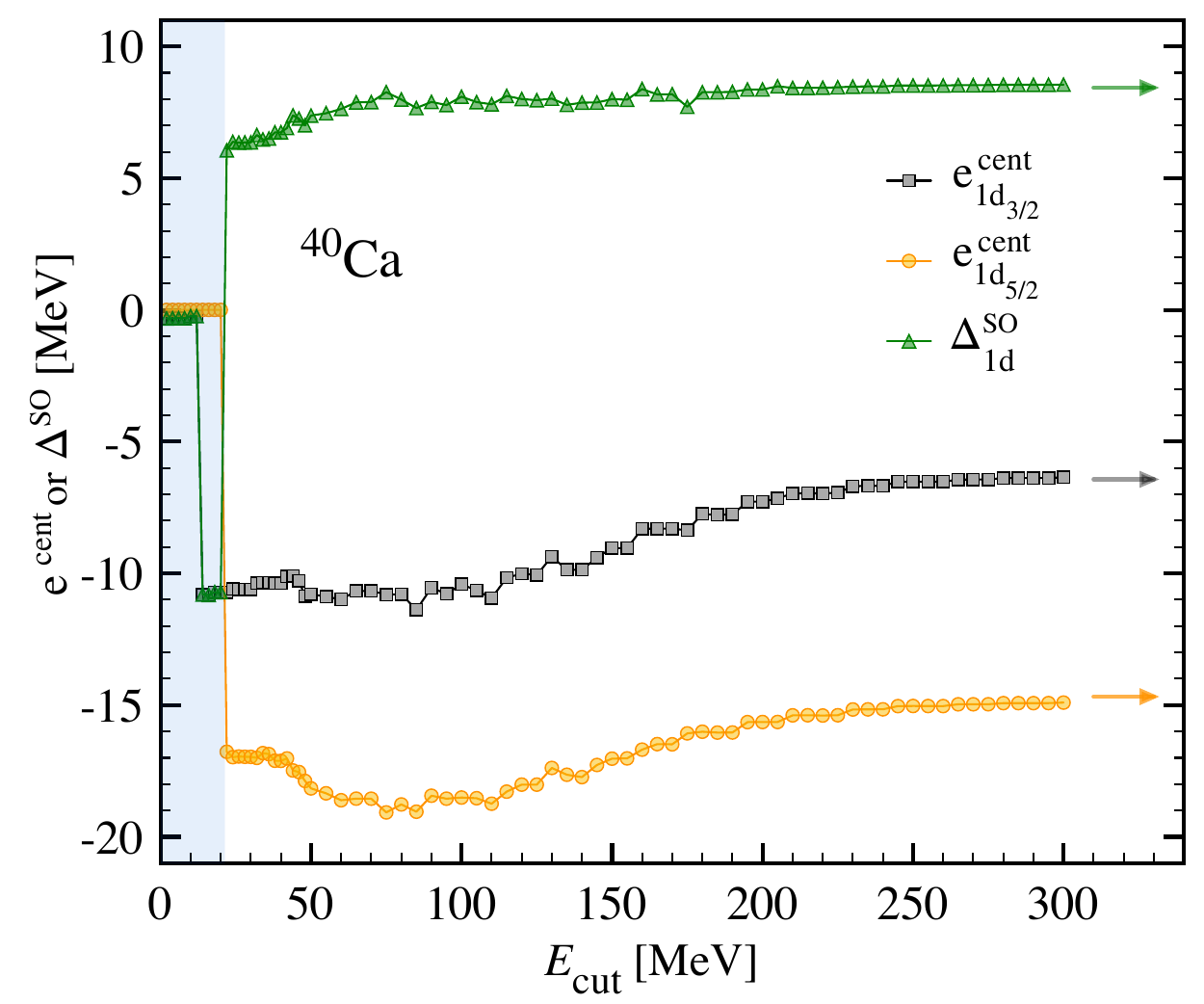}
\caption{Neutron $1d_{3/2}$ and $1d_{5/2}$ ESPEs in $^{36}$S (\emph{left}) and $^{40}$Ca (\emph{right}) as a function of an energy cutoff $E_\text{cut}$ used to truncate both sums entering Eq.~\eqref{defsumrule} according to $|E^{\pm}_\lambda|\leq E_\text{cut}$. Their difference ($\Delta^\text{SO}_{1d}$) is also shown. Shaded areas represent energy cuts excluding the main $3/2^+$ and $5/2^+$ fragments. Arrows on the right point to values computed without energy cutoffs, i.e. to the full ESPEs and $\Delta^\text{SO}_{1d}$. Calculations are performed via GGF-ADC(2) with the \sat{} Hamiltonian.
}
\label{fig_diagS}
\end{figure*}

The \emph{increase} of  $\Delta^\text{SO}_{1d}$ going from $^{36}\text{S}$ to $^{40}\text{Ca}$ amounts to $2.2$ MeV, at variance with the \emph{decrease} of about 500 keV predicted by Mairle's trend~\cite{Mairle93}.
The latter trend is understood within an independent-particle picture as being driven by the one-body mean-field spin-orbit potential originating from the two-body spin-orbit interaction entering the nuclear Hamiltonian.
The increase that is presently observed has been instead interpreted as being due to the action of the one-body mean-field spin-orbit potential originating from the two-body tensor interaction, which is expected to enhance the neutron $\Delta^\text{SO}_{1d}$ as protons fill the $1d_{3/2}$ shell~\cite{Otsuka20}. Present ab initio structure calculations based on Hamiltonians naturally including tensor interactions support this interpretation~\cite{Jongile23}.

\subsection{Stability of ESPEs}
\label{sec_sensitivity}

Having obtained ab initio GGF $e^{\text{cent}}_{1d_{3/2}}$ and $e^{\text{cent}}_{1d_{5/2}}$, along with the associated neutron spin-orbit splitting  $\Delta^\text{SO}_{1d}$, the goals of the present section are (i) to gauge the impact of making two approximations to their computation and (ii) to assess theoretical uncertainties stemming from the ab initio scheme.

\subsubsection{Energy cutoff}
\label{cutoffE}

As mentioned earlier, the direct-reaction approach to ESPEs is restricted to the energy range accessible via one-neutron addition and/or removal experiments. For example, the  impressive $^{36}\text{S}(p,d)^{35}\text{S}$ reaction relevant to the physical case of present interest could reach 98 final states in $^{35}\text{S}$ up to an excitation energy of 16 MeV~\cite{Jongile23}. 
Arguments that the earlier $(p,d)$ experiment $^{40}\text{Ca}$ limited to about $9.5$\,MeV excitation energy~\cite{Matoba93} did not miss states with significant cross sections were made in order to ensure that $e^{\text{cent}}_{1d_{3/2}}$ and $e^{\text{cent}}_{1d_{5/2}}$ calculated in both nuclei could be safely compared. Still, it is difficult to determine to what extent the missing strength affects computed ESPEs. 

Using ab initio GGF-ADC(2), the impact of reducing the accessible energy range is now illustrated. To do so, the one-neutron addition and removal channels are simultaneously truncated according to $|E^{\pm}_\lambda|\leq E_\text{cut}$ in the construction of the Baranger Hamiltonian (Eq.~\eqref{defsumrule}). While the lessons learnt cannot be transposed quantitatively to the direct-reaction approach, the idea is to gauge the impact of omitting many high-energy final states carrying small spectroscopic strengths.

Figure~\ref{fig_diagS} displays the results in $^{36}$S and $^{40}$Ca. Increasing $E_\text{cut}$ , $e^{\text{cent}}_{1d_{3/2}}$ and $e^{\text{cent}}_{1d_{5/2}}$ strongly dive down in both nuclei as the dominant low-lying states are included in the calculation of the centroid matrix $\mathbf{h}$. In $^{36}$S, the significant fragmentation of the $5/2^+$ strength makes $e^{\text{cent}}_{1d_{5/2}}$ readjust several times in the first $10-20$\,MeV interval. Contrarily, the single dominant peak corresponding to the $3/2^+$ ground state of $^{35}$S already settles $e^{\text{cent}}_{1d_{3/2}}$ at about $E_\text{cut}\approx 10$\,MeV. In $^{40}$Ca, the fingerprint of the unfragmented $5/2^+$ strength is testified by a single jump around 20\,MeV. 
Eventually, the shaded areas denote the minimal energy range (i.e. $E_\text{cut}\approx 25$\, MeV) to be covered in order to incorporate the main $3/2^+$ and $5/2^+$ fragments in both nuclei. Using a smaller value of $E_\text{cut}$ is empirically unreasonable as already discussed above. 
Consequently, the effect of further increasing the accessible energy range must be analyzed with respect to the values obtained for $E_\text{cut}\approx 25$\, MeV that are thus used as a baseline. 
As a matter of fact, $E_\text{cut}\approx 25-30$\, MeV is representative of the energy range covered by the $(p,d)$ reactions performed in connection with the physical case of present interest~\cite{Jongile23,Matoba93}. 
Similarly, the valence-space shell model also employed to interpret the evolution of the spin-orbit splitting computed final states up to about $20$\,MeV excitation energy, i.e. up to $E_\text{cut} \approx 30$\,MeV.

Beyond the shaded area, one observes that the large number of states carrying (very) small spectroscopic strength\footnote{Such small strengths are not visible when using a linear scale as was done in Figs.~\ref{fig_strengthS} and \ref{fig_strengthCa}.} accumulate to impact ESPEs significantly. Typically, $e^{\text{cent}}_{1d_{5/2}}$ ($e^{\text{cent}}_{1d_{3/2}}$) is first driven down in both nuclei by about $2$-$3$\,MeV ($0$-$1$\,MeV) over the interval $E_\text{cut} \in [25,75]$\,MeV before going up by about $4$\,MeV ($4$\,MeV) and stabilizing at $E_\text{cut} \approx 300$\,MeV. The latter value is characteristics of the energy scale imprinted in the $\chi$EFT Hamiltonian employed in the ab initio theoretical scheme and is thus itself a marker of the employed theoretical scheme.

Eventually, presently studied ESPEs increase by a net value of a few MeV (between 1 and 5) when going from $E_\text{cut}\approx 25$\,MeV to $E_\text{cut} \approx 300$\,MeV while varying non-trivially in between. This non-trivial evolution is however similar for $e^{\text{cent}}_{1d_{5/2}}$ and $e^{\text{cent}}_{1d_{3/2}}$. This results into a spin-orbit splitting $\Delta^\text{SO}_{1d}$ that varies much less for $E_\text{cut} \in [25, 300]$\,MeV. Still, $\Delta^\text{SO}_{1d}$ increases by about $1$\,MeV in $^{36}$S and by about $2.5$\,MeV in $^{40}$Ca, eventually augmenting significantly the {\it change} of that spin-orbit splitting when going from $^{36}$S to $^{40}$Ca.

\subsubsection{Diagonal approximation}
\label{weightsum}

In the previous section, the impact of reducing the accessible energy range was investigated {\it while} maintaining the non-trivial diagonalisation of the Baranger Hamiltonian at play within the ab initio scheme. However, a further hypothesis built into the direct-reaction and the valence-space shell model approaches relates to the use of a single harmonic oscillator basis state per angular momentum.
While being absent from the ab initio scheme, such a feature can be enforced as an {\it approximation} by omitting all off-diagonal elements of the spectroscopic probability matrices (Eq.~\eqref{spectroproba}) expressed in the spherical harmonic oscillator basis. This corresponds to replacing the Baranger eigenvalue problem (Eq.~\eqref{HFfield3}) with the weighted sum (Eq.~\eqref{HFfield2}) 
\begin{eqnarray}
e^{\text{cent}}_{p} &\approx&  \sum_{\mu \in {\cal H}_{A\!+\!1}} S_{\mu}^{+pp} E_{\mu}^{+} + \sum_{\nu\in {\cal H}_{A\!-\!1}}  S_{\nu}^{-pp} E_{\nu}^{-}  \,\,\,. \label{HFfield2Appr}
\end{eqnarray}
computed directly in the spherical harmonic oscillator basis $\{a^\dagger_{p}\}$\footnote{This approximation leads thus to omitting the difference between the Baranger basis and the initial harmonic oscillator basis induced by many-body correlations.}. Such an approximation can be further combined with the reduction of the accessible energy range introduced in the previous section. 

\begin{figure}
\centering
\includegraphics[width=1.0\columnwidth]{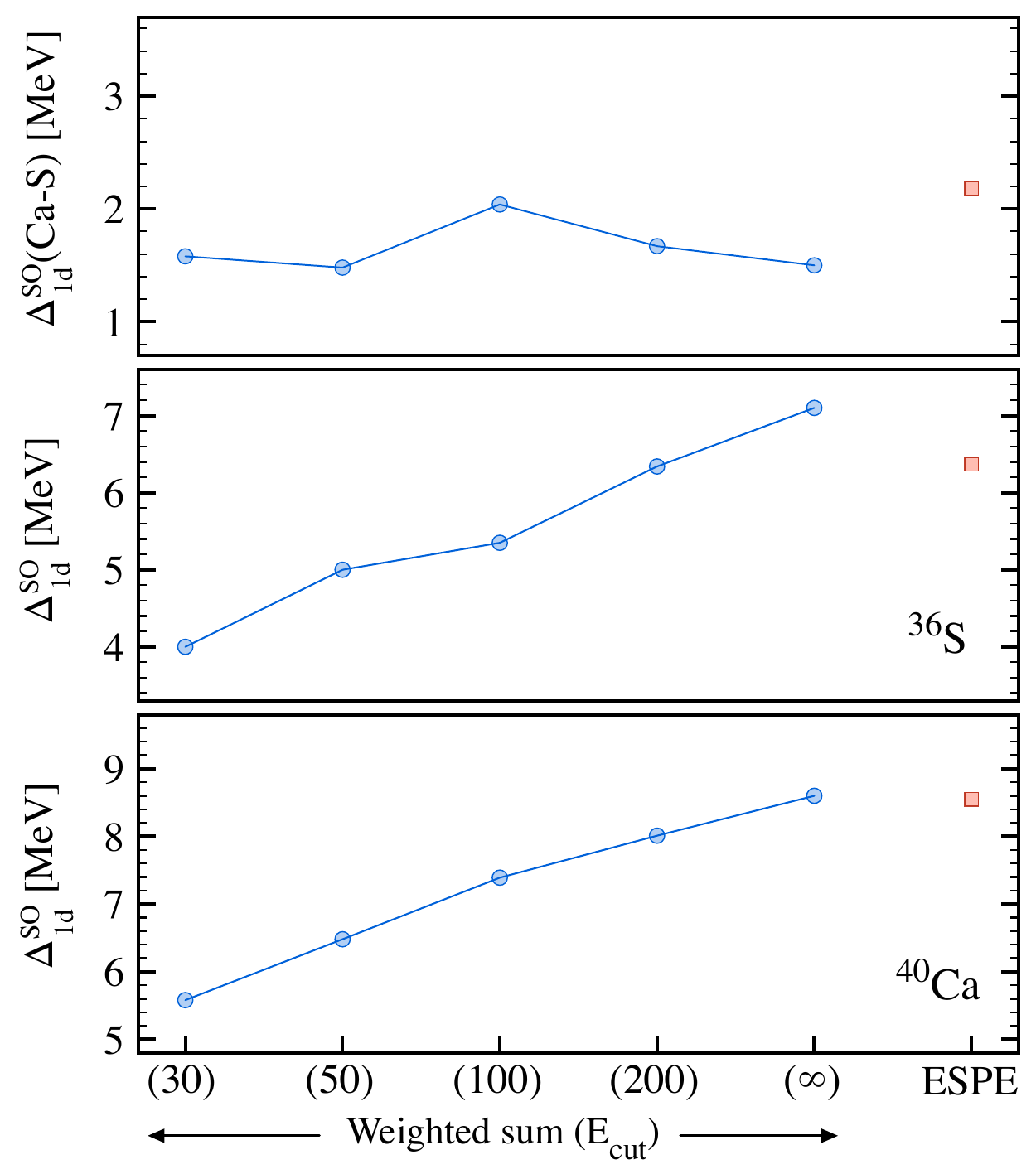}
\caption{(\emph{Bottom}) GGF-ADC(2) one-body spin-orbit splitting $\Delta^\text{SO}_{1d}$ in $^{40}$Ca computed via Eq.~\eqref{HFfield2Appr} with different values of $E_\text{cut}$. The spin-orbit splitting computed from the full Baranger ESPEs is also shown for reference on the right (red square).
(\emph{Middle}) Same as in the bottom panel for $^{36}$S.
(\emph{Top}) Variation of the spin-orbit splitting from $^{36}$S to $^{40}$Ca  obtained by taking the difference between the points in the bottom and the middle panels.
Calculations are performed with the \sat{} Hamiltonian.
}
\label{fig_compESPE}
\end{figure}

Ab initio GGF-ADC(2) results obtained from Eq.~\eqref{HFfield2Appr} for different values of $E_\text{cut}$ are displayed in Fig.~\ref{fig_compESPE}. The right-most points (red squares) labelled as ``ESPE" are reference values obtained without any approximation. In the bottom (middle) panel the spin-orbit splitting $\Delta^\text{SO}_{1d}$ in $^{40}$Ca ($^{36}$S) is shown. The top panel displays their difference, i.e. the evolution of the one-body spin-orbit splitting between $^{36}$S and $^{40}$Ca, which in the full calculation amounts to about $2.2$\,MeV. The second set of points, coming from the right, reports results from the weighted sum~\eqref{HFfield2Appr} computed in the original spherical harmonic oscillator basis without applying any cut on the accessible energy range. While $\Delta^\text{SO}_{1d}$ increases by $0.1$\,MeV in $^{40}$Ca, it is increased by $0.8$\,MeV in $^{36}$S. This results into a non-negligible reduction of the $2.2$\, MeV increase of $\Delta^\text{SO}_{1d}$ between $^{36}$S and $^{40}$Ca by about $700$\,keV. This change is driven by $e^{\text{cent}}_{1d_{5/2}}$ in $^{36}$S, i.e. the more fragmented the diagonal strength, the larger the associated off-diagonal matrix elements of the spectroscopic probability matrices, and thus the greater the impact of the present approximation. Going further left in the figure adds the effect of reducing the accessible energy range. Consistently with the result shown in the previous section, this reduction decreases $\Delta^\text{SO}_{1d}$ in both nuclei down to $E_\text{cut} = 30$\,MeV. However, the reduction is presently larger in $^{36}$S due to the more important impact of the second approximation on $e^{\text{cent}}_{1d_{5/2}}$ in that nucleus. In the end the combination of {\it both} approximations makes the change even more similar in the two nuclei than in the previous section such that the impact on the {\it evolution} of $\Delta^\text{SO}_{1d}$ is actually {\it reduced}. Eventually, while $\Delta^\text{SO}_{1d}$ changes by $2.2$\,MeV when using full Baranger ESPEs, it presently reaches $1.6$\,MeV at $E_\text{cut} = 30$\,MeV. 

It interesting to observe that imposing two approximations (i.e. diagonality of the strength function and energy range limited to $E_\text{cut} \approx 30$\,MeV) consistent with hypotheses built into the valence-space shell model scheme makes the change of $\Delta^\text{SO}_{1d}$ computed within the ab initio scheme more consistent with the value obtained with the former approach~\cite{Jongile23}\footnote{And closer to the even smaller value obtained within the direct-reaction approach characterized by an even greater reduction of $E_\text{cut}$~\cite{Jongile23}.}. Even though it must be taken with a grain of salt because imposing a posteriori an approximation to a given theoretical scheme is obviously different from building it a priori into a theoretical scheme\footnote{Furthermore, the practical way that is presently used to reduce the accessible energy range within the ab initio scheme is only remotely related to the way such a limitation is built into the shell model calculation.}, this observation qualitatively establishes an understanding of the scheme dependence of the ESPEs obtained within the two theoretical schemes. Eventually, none of the two is more ``real/correct" than the other but at least the different results (and interpretations) can be rationalised.
\begin{figure}
\centering
\includegraphics[width=1.0\columnwidth]{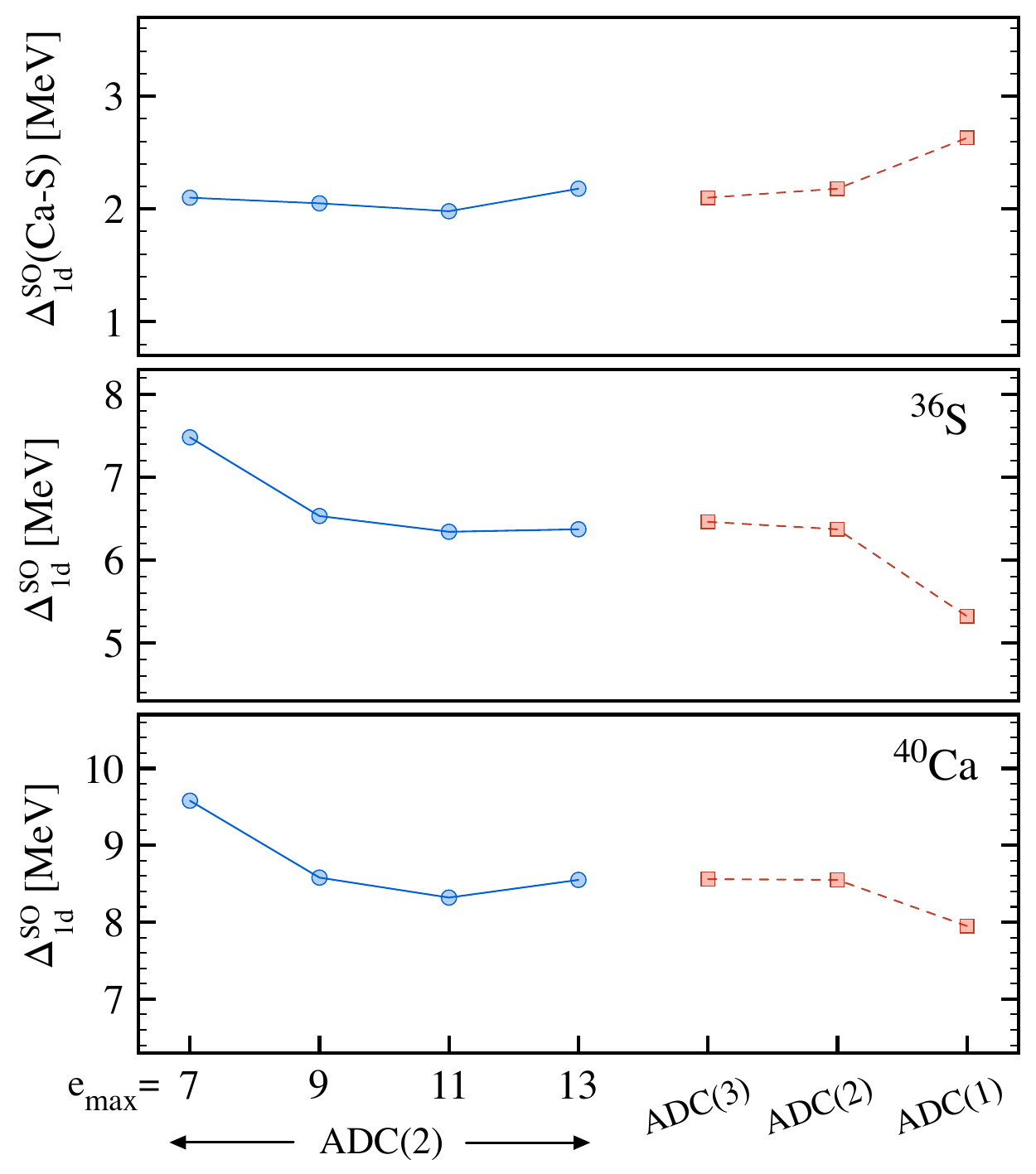}
\caption{(\emph{Bottom}) One-body spin-orbit splitting $\Delta^\text{SO}_{1d}$ in $^{40}$Ca computed from Baranger ESPEs. On the left, GGF-ADC(2) results for increasing values of $e_\text{max}$ are shown. On the right, $\Delta^\text{SO}_{1d}$ is displayed at different levels of ADC truncation.
(\emph{Middle}) Same as in the bottom panel for $^{36}$S.
(\emph{Top}) Variation of the spin-orbit splitting from $^{36}$S to $^{40}$Ca  obtained by taking the difference between the points in the bottom and the middle panels.
Calculations are performed with the \sat{} Hamiltonian.
}
\label{fig_compESPE_emax}
\end{figure}
\begin{figure}
\centering
\includegraphics[width=1.0\columnwidth]{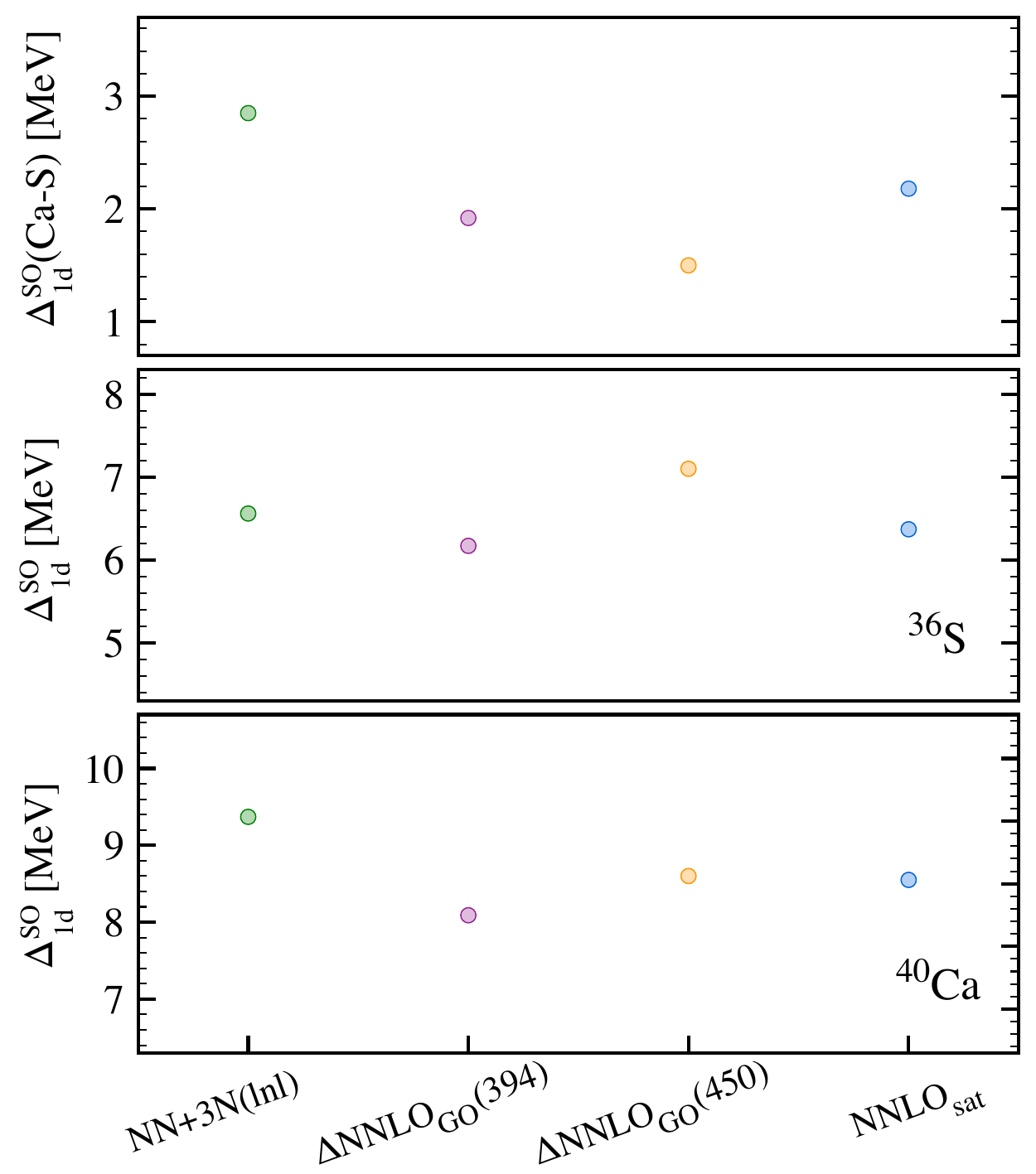}
\caption{(\emph{Bottom}) GGF-ADC(2) one-body spin-orbit splitting $\Delta^\text{SO}_{1d}$ in $^{40}$Ca computed from Baranger ESPEs with different input Hamiltonians.
(\emph{Middle}) Same as in the bottom panel for $^{36}$S.
(\emph{Top}) Variation of the spin-orbit splitting from $^{36}$S to $^{40}$Ca  obtained by taking the difference between the points in the bottom and the middle panels.
}
\label{fig_compESPE_int}
\end{figure}

\subsubsection{Theoretical uncertainties}
\label{errors}

Different sources of theoretical error characterise the present computational scheme and thus potentially affect the above analysis.
Specifically, uncertainties of ab initio calculations can be separated in three categories: (i) basis truncation, (ii) many-body truncation, (iii) Hamiltonian truncation.
The first two are addressed in Fig.~\ref{fig_compESPE_emax}, where $\Delta^\text{SO}_{1d}$ in the two nuclei is first shown for increasing values of $e_\text{max}$.
One notices a similar convergence in the two cases, with differences of only few hundred keV between $e_\text{max}=9$ and $e_\text{max}=13$. 
As a result, the change in spin-orbit splitting, shown in the top panel of Fig.~\ref{fig_compESPE_emax}, remains very stable against basis truncation\footnote{The impact of the three-body truncation $e_\text{3max}$ has also been assessed and found to be negligible.}.
Within the present context, the impact of approximations on the solution of the A-body Schr\"{o}dinger equation can be investigated by performing calculations at different ADC truncation levels, namely ADC(1), ADC(2) and ADC(3).
The resulting values of $\Delta^\text{SO}_{1d}$ are also displayed in Fig.~\ref{fig_compESPE_emax}.
Similarly to the model truncation, one observes a clear convergence when the truncation order is increased, the difference between ADC(2) and ADC(3) amounting to less than 100 keV in all cases.
The analysis summarised in Fig.~\ref{fig_compESPE_emax} has been repeated with the \lnl interactions, yielding analogous results.
These variations are much smaller than the corresponding ones on e.g. total energies or even single excitation energies (see, e.g. Refs.~\cite{Soma20a, Soma21}), suggesting that these uncertainties cancel out to a large extent when computing quantities like $\Delta^\text{SO}_{1d}$.
One thus concludes that basis truncation and many-body truncation do not affect significantly the stability of spin-orbit splittings in these nuclei.

To estimate the corresponding impact of the input Hamiltonian, calculations with a few different interactions have been performed\footnote{This represents a poor man's version of both a more systematic order-by-order variation of Weinberg's chiral EFT expansion and a more rigorous study of the dependence on the SRG scale, as done e.g. in Refs.~\cite{Huther19, Porro24a}. In the  context of ESPEs, the latter dependence was investigated in detail in Ref.~\cite{Duguet15b}.}.
The resulting values of $\Delta^\text{SO}_{1d}$ for the two nuclei, as well as their difference, are shown in Fig.~\ref{fig_compESPE_int}.
A spread of about 1 (1.5) MeV is observed for $^{36}$S ($^{40}$Ca), the value obtained with \sat{} being located in the middle of the interval.
The slightly larger variation associated to $^{40}$Ca calculations reflects the difficulty to correctly describe $5/2^+$ separation energies and fragmentation at the GGF-ADC(2) level in this nucleus. 
These variations do not fully cancel out in the difference of the two splittings, which displays a spread of about 1.3 MeV.
While subleading for the single $\Delta^\text{SO}_{1d}$ (e.g. compared to the approximations studied in Figs.~\ref{fig_diagS} and \ref{fig_compESPE}), the uncertainty associated with the nuclear Hamiltonian thus becomes quantitatively relevant when considering the variation of the spin-orbit splitting. 
Nevertheless, just as for the approximations addressed in the previous sections, such an uncertainty does not qualitatively affect the physical picture, i.e. does not change the fact that $\Delta^\text{SO}_{1d}$ undergoes a sizeable reduction from calcium to sulfur.

\section{Conclusions}
\label{sec_conclusions}

The notion of shell structure based on the computation of effective single-particle energies (and associated one-body basis) relies on a well-defined and non-ambiguous procedure. However, the outcome of the procedure does depend on the particular theoretical scheme and scale under consideration, i.e. the choice of dynamical degrees of freedom and their interactions along with the implicit fixation of the unitary freedom at play in the associated quantum mechanics.

The present work of pedagogical character illustrated the calculation and use of effective single-particle energies. 
This was done through a specific application of the GGF ab initio theoretical scheme to the evolution of the splitting between neutron $\ell=2$ spin-orbit partners along $\text{N}=20$ isotones.
After displaying the picture extracted from the ab initio scheme, two features entering the computation of ESPEs were altered to empirically exemplify the differences that can be expected from ESPEs computed within alternative theoretical schemes, e.g. the valence-space shell model, in which such alterations are valid by construction, or the direct-reaction approach based on one-nucleon transfer data.
Doing so, the increase of the neutron $1d_{3/2}$-$1d_{5/2}$ ab initio one-body spin-orbit splitting obtained when going from $^{36}$S to $^{40}$Ca could become consistent with the ones computed within the valence-space shell model and the direct-reaction framework.
The theoretical uncertainties associated with the calculation of $\Delta^\text{SO}_{1d}$ do not significantly affect this picture.

This consistency must be taken with a grain of salt because imposing a posteriori an approximation to a given theoretical scheme is clearly different from building it a priori.
Nevertheless, this observation qualitatively establishes an understanding of the scheme dependence of ESPEs and thus illustrates the possible degree of variation in the non-observable shell structure emerging from the correlated many-body wave function.
Eventually, none of the realisable interpretations in terms of ESPEs is more real or correct than the others but at least the differences obtained within the various schemes can be rationalised.
 
\section*{Acknowledgments}

The authors wish to thank J.-P.~Ebran, S.~Jongile, R.~Neveling, O.~Sorlin, M.~Wiedeking and C.~Yuan for useful discussions, as well as A.~Scalesi for providing several sets of interation matrix elements.
Calculations were performed by using HPC resources from GENCI-TGCC (Contract No. A0150513012).

\bibliography{newbiblio}

\end{document}